\RequirePackage{fix-cm}
\documentclass[smallextended]{svjour3}       
\smartqed  
\usepackage[utf8]{inputenc}
\usepackage{amsmath}
\usepackage{amssymb}

\usepackage{listings}
\usepackage{color}
\usepackage{textcomp}
\usepackage{algpseudocode}
\usepackage{algorithm}

\usepackage{varwidth}
\usepackage{xcolor}
\usepackage{enumitem}
\newcommand{\subscript}[2]{$#1 _ #2$}
\usepackage{graphicx}
\usepackage{subfig}

\usepackage{xspace}
\newcommand{\projectName}{\emph{\'{O}bidos}\xspace}

\definecolor{mygreen0}{rgb}{0, 0.75, 0}

\definecolor{myred1}{rgb}{1,0,0}
\definecolor{mygreen1}{rgb}{0, 1, 0}
\definecolor{myblue0}{rgb}{0, 0, 1}

\definecolor{myred2}{rgb}{1,0.5,0.5}
\definecolor{mygreen2}{rgb}{0.5, 1, 0.5}
\definecolor{myblue2}{rgb}{0.5, 0.5, 1}

\definecolor{mygreen}{rgb}{0, 0.25, 0}
\definecolor{myblue}{rgb}{0, 0, 0.75}
\definecolor{myred0}{rgb}{0.5,0,0}

\definecolor{listinggray}{gray}{0.98}
\definecolor{lbcolor}{rgb}{0.98,0.98,0.98}
\algdef{SE}[VARIABLES]{Variables}{EndVariables}
   {\algorithmicvariables}
   {\algorithmicend\ \algorithmicvariables}
\algnewcommand{\algorithmicvariables}{\textbf{global variables}}

\algdef{SE}[LARIABLES]{Lariables}{EndLariables}
   {\algorithmiclariables}
   {\algorithmicend\ \algorithmiclariables}
\algnewcommand{\algorithmiclariables}{\textbf{local variables}}

\begin{document}

\pagenumbering{gobble}
\large
\vfil
\begin{center}
\begin{minipage}{18cm}
\end{minipage}
\end{center}
\vfil
\includegraphics[width=0.4\columnwidth]{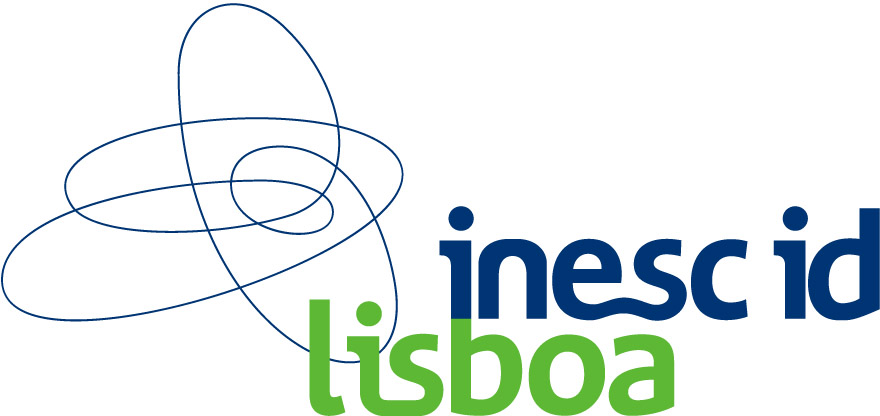}
\vfil
\begin{center}
\vskip 2cm
{\LARGE On-Demand Big Data Integration}\\
{\large A Hybrid ETL Approach for Reproducible Scientific Research\\}
\vskip 1cm

{\small Pre-print Submitted to the DMAH Special Issue of the Springer DAPD Journal}

\vskip 1cm
{\Large --- INESC-ID Lisboa Tech. Rep. 3/2018 ---}

\vskip2cm

{\Large Pradeeban Kathiravelu$^{1,2}$, Ashish Sharma$^{3}$,\\
Helena Galhardas$^{1}$, Peter Van Roy$^{2}$, Lu{\'\i}s Veiga$^{1}$}\\
\vskip 3mm
{\small $^{1}$INESC-ID / Instituto Superior Técnico, Universidade de Lisboa, Portugal,\\$^{2}$Université catholique de Louvain, Belgium, $^{3}$Emory University, GA, USA}
\end{center}
\vfil
\begin{center}
\begin{minipage}{18cm}
\end{minipage}
\end{center}
\vfil
\begin{center}
April 2018
\end{center}
\normalsize

\cleardoublepage
\pagenumbering{arabic}

\title{On-Demand Big Data Integration}
\subtitle{A Hybrid ETL Approach for Reproducible Scientific Research}

\author{Pradeeban Kathiravelu  \and \\ Ashish Sharma \and Helena Galhardas \and \\ Peter Van Roy \and Lu{\'\i}s Veiga
}

\institute{P. Kathiravelu \at
            INESC-ID Lisboa/Instituto Superior Técnico, Universidade de Lisboa, Lisboa, Portugal \\
Université catholique de Louvain, Louvain-la-Neuve, Belgium\\
              \email{pradeeban.kathiravelu@tecnico.ulisboa.pt}   
           \and
           A. Sharma \at
            Emory University School of Medicine, Atlanta, USA \\
              \email{ashish.sharma@emory.edu }
                         \and
           H. Galhardas \at
            INESC-ID Lisboa/Instituto Superior Técnico, Universidade de Lisboa, Lisboa, Portugal \\
              \email{helena.galhardas@tecnico.ulisboa.pt}
           \and
           P. Van Roy \at
            Université catholique de Louvain, Louvain-la-Neuve, Belgium \\
              \email{peter.vanroy@uclouvain.be }
           \and
           L. Veiga \at
            INESC-ID Lisboa/Instituto Superior Técnico, Universidade de Lisboa, Lisboa, Portugal \\
              \email{luis.veiga@inesc-id.pt}
}
\date{Received: date / Accepted: date}

\maketitle
\begin{abstract}
Scientific research requires access, analysis, and sharing of data that is distributed across various heterogeneous data sources at the scale of the  Internet. An eager ETL process constructs an integrated data repository as its first step, integrating and loading data in its entirety from the data sources. The bootstrapping of this process is not efficient for scientific research that requires access to data from very large and typically numerous distributed data sources. a lazy ETL process loads only the metadata, but still eagerly. Lazy ETL is faster in bootstrapping. However, queries on the integrated data repository of eager ETL perform faster, due to the availability of the entire data beforehand. 

In this paper, we propose a novel ETL approach for scientific data integration, as a hybrid of eager and lazy ETL approaches, and applied both to data as well as metadata. This way, Hybrid ETL supports incremental integration and loading of metadata and data from the data sources. We incorporate a human-in-the-loop approach, to enhance the hybrid ETL, with selective data integration driven by the user queries and sharing of integrated data between users. We implement our hybrid ETL approach in a prototype platform, \projectName, and evaluate it in the context of data sharing for medical research. \projectName outperforms both the eager ETL and lazy ETL approaches, for scientific research data integration and sharing, through its selective loading of data and metadata, while storing the integrated data in a scalable integrated data repository.


\end{abstract}

\section{Introduction}
\label{sec:intro}
Big data integration is crucial for numerous application domains, such as reproducible science~\cite{reichman2011challenges}, medical research~\cite{lee2009knowledge}, and transport planning~\cite{huang2003data}, to enable data analysis and information retrieval. Scientific research often requires access to big data from various data sources, often geographically distributed~\cite{hey2005cyberinfrastructure}. Scientific data is typically heterogeneous, including binary and textual data, and stored in structured, semi-structured, or unstructured formats. In addition, data sources usually support distinct data access interfaces, ranging from database SQL queries to service-based APIs (Application Programming Interfaces)~\cite{heinzlreiter2014cloud}. Effectively and efficiently integrating such diversity and quantity of data is challenging.

To discover compelling scientific insights from data, it is often required to extract, transform, and load it into an \textit{integrated data repository} (e.g., a data warehouse~\cite{chaudhuri1997overview}). This process is typically called ETL (Extract, Transform, and Load)~\cite{vassiliadis2009survey}. An ETL process makes data accessible through a uniform schema, by constructing an integrated data repository. Thus it supports fast and efficient querying of the scientific research data.

\paragraph*{\textbf{ETL Efficiency:}}~
Traditionally, ETL has been an eager process, loading the entire content of the data sources into an integrated data repository as a first step. However, {\it eager ETL} is often unsuitable for handling scientific data. First, the bootstrapping process of eager data integration and loading takes too long. This time waste is unnecessary for scientific research~\cite{ccaparlar2016scientific} that often requires only a subset of data. Second, entirely integrating and loading the contents of data sources can be challenging due to the substantial resource demands. In fact, it requires high loading time and bandwidth. Furthermore, eager ETL also demands large storage due to the typical amount of data to integrate. Third, scientific data sources are often accessible only to authorized people. Loading the entire contents of data sources into an integrated data repository may enable to bypass the data authorization permissions established for data sources. Users would then be able to access data from the integrated data repository, thus increasing the probability of data access violation.

{\it Lazy ETL}~\cite{kargin2013lazy} aims at mitigating the limitations of eager ETL, by integrating and loading the data only when necessary. Concretely, it avoids loading the entire contents of data sources into an integrated data repository as the initial step. A data source is composed of several data entries. For binary data, there is typically a piece of textual metadata (containing identifying information) attached to each data entry, in the file header. Metadata is often sufficient for the initial scientific research demands. As an illustrative example, consider the medical images stored in DICOM (Digital Imaging and Communications in Medicine)~\cite{mildenberger2002introduction} standard format in various data sources such as the Cancer Imaging Archive (TCIA)~\cite{clark2013cancer}. The DICOM image file is in binary format. Often, there is textual metadata associated with each image. The DICOM metadata includes the image identification constituted by the series, study, and the identification of the patient the image belongs to. The metadata can be leveraged in the early stages of medical research, while DICOM image processing can be performed at a later phase, only for images selected as relevant (from the metadata). Thus, lazy ETL advocates for eagerly integrating and loading only the metadata, instead of the data entry itself (that is addressed lazily). Integrating and loading the metadata, in this case, is faster than loading the entire data entry, due to the substantially smaller size of the metadata. Therefore, lazy ETL usually bootstraps faster than eager ETL.

Scientific experiments are often repeated several times by multiple researchers to confirm the accuracy of the outcomes. Therefore, frequent and repetitive queries are common. As a consequence, persistently storing the previously processed data entries into the integrated data repository would make recurring scientific research experiments faster. While eager ETL loads the data entirely into an integrated data repository, current lazy ETL approaches are not able to persistently store data required for previous queries. This means that recurring scientific queries execute slower under lazy ETL than under eager ETL. The gain obtained by faster data integration and loading in lazy ETL is lost when executing recurring queries because they cannot use stored results from previous queries.

\paragraph*{\textbf{Scalability:}}~ Scientific research often requires integrating large amounts of heterogeneous data from several web data sources~\cite{dong2013big}. Consequently, even an eager metadata-only ETL process (as prescribed by lazy ETL) can be challenging in scientific research, due to the distributed and heterogeneous nature of data sources. Moreover, metadata of some data sources tend to be as large as or larger than the data entries themselves. For example, Scality RING petascale object storage~\cite{scality1} consists of metadata up to 10 times larger than the data entries, supporting content-based searches through its metadata (designed for indexing). A typical lazy ETL process may fail to outperform an eager ETL process in bootstrapping in the presence of such data sources, due to the large size of metadata.

In practice, the researcher is often aware of the specific datasets that she needs and the characteristics of the data sources those datasets belong to. So, the researcher may be able to directly access the required data without accessing and querying the corresponding metadata. For example, consider a research study comparing the effects of an experimental medicine against with those of a placebo for variants of brain tumor. For this research study, the researcher only needs to load the imaging data of brain tumor from the data sources. Moreover, the researcher often possesses insights of the data such as the location of relevant image collections and the type of data access that is provided. Therefore, she can directly query the data sources and then load only specific subsets of the metadata, rather than eagerly loading the whole metadata.

Additionally, since the number of web data sources, as well as the amount of data and metadata, tend to increase, the storage requirements for the integrated data repository must be adaptable. In particular, a scalable storage is essential to accommodate data and metadata selectively accessed and incrementally integrated and loaded by the researcher. However, the current ETL approaches do not support such a selective ETL process into a scalable integrated data repository.

\paragraph*{\textbf{Interoperability and Human Intervention:}}~
Extracting and transforming data from web sources must consider various data storage and access interfaces. Data storage formats and access interfaces have been standardized in various research fields, to facilitate seamless access to the heterogeneous data sources. For example, Health Level Seven International (HL7) Fast Healthcare Interoperability Resources (FHIR)~\cite{fhir} is a standard for consistent data exchange between healthcare applications. Despite the popularity of these standards, a vast majority of data sources still fail to adhere to them. Thus, interoperability between heterogeneous data sources is still lacking~\cite{kadadi2014challenges}. Consequently, data integration across various scientific web data sources is challenging, and typically not possible in an effective and efficient way without human involvement.

Currently, in some domains, ETL is performed on-demand by a user~\cite{krishnan2016towards}. The user is involved in the ETL process by incrementally integrating and loading subsets of data or metadata that are relevant to a given research question. The user is often aware of the details about data source access and data location. This expert knowledge could and should be incorporated into the ETL process. This type of user involvement is called \textit{human-in-the-loop ETL}. It often consists of two parts. First, the user manually searches and downloads the datasets from the web data sources. Then, she integrates and stores the result in an integrated data repository. By narrowing down the search space to a smaller subset of relevant data sources, human-in-the-loop ETL shortens the data integration and loading time. However, existing ETL frameworks do not support the automatic incorporation of human in the process. Therefore, currently, human-in-the-loop ETL process remains a cumbersome manual and repetitive task.

\paragraph*{\textbf{Efficient Scientific Data Sharing:}}~
Data used in a scientific research study often needs to be shared among researchers for collaboration and reproducibility purposes. However, this process is not efficient. First, sharing data by replicating its contents creates an excessive overhead on bandwidth, storage, and data maintenance. Therefore, data must be shared with minimal data replication.
Second, researchers interested in the data resulting from an integration process may belong to a single organization. Nevertheless, collaboration can extend beyond the organizational boundaries, but the repetition of the same ETL process to obtain the same integrated data must be avoided.
Current ETL approaches do not consider data sharing of the integrated data repositories beyond the organization. Therefore, the integrated data are often manually shared, in an approach oblivious to the ETL process. Such data sharing is inefficient and may lead to the existence and maintenance of duplicate data.

\paragraph*{\textbf{Motivation:}}~
Given the above premises, we aim at addressing the following research questions in this paper: 
\begin{enumerate}[label=(\subscript{RQ}{{\arabic*}})]
\setlength{\itemindent}{1.8em}
\item Can we increase the speed of the bootstrapping process in ETL by selectively accessing, integrating, and loading metadata?
\item Can we achieve faster execution time for repetitive scientific research queries by storing the previously integrated and loaded data in an integrated data repository?
\item Can we incorporate the human knowledge into an ETL framework to selectively and incrementally integrate and load only the relevant subsets of metadata or data, from web data sources?
\item Can the relevant subsets of data and metadata loaded by a research scientist for a specific experiment be shared efficiently for reproducibility purposes, thus minimizing data replication across peers from multiple organizations and avoiding the repetition of the ETL process?
\end{enumerate}

\paragraph*{\textbf{Contributions:}}~
The goal of this paper is to answer the identified research questions, focusing on medical research as motivating real-life domain. The main contributions of this paper are:

\begin{enumerate}
        \setlength{\itemindent}{1.8em}
\item A novel hybrid ETL approach for accessing and integrating data and metadata from heterogeneous data sources, and loading the resulting data into a scalable integrated data repository. ($RQ_1$ and $RQ_2$)
\item The incorporation of human knowledge into a hybrid ETL process to selectively integrate and load subsets of data and metadata on-demand. ($RQ_3$)
\item A data sharing mechanism that enables to virtually share the relevant datasets efficiently through ``pointers'' to data, instead of repeatedly loading and replicating the actual data and metadata. ($RQ_4$)
\end{enumerate}

We implemented \projectName\footnote{\projectName is a medieval fortified town that has been patronized by various Portuguese queens. It is known for its sweet wine, served in a chocolate cup.}, an on-demand big data integration platform for scientific research. We presented a preliminary version of \projectName in our previous work~\cite{kathiravelu2017demand}. In this paper, we elaborate in detail, how \projectName supports hybrid ETL enhanced with human-in-the-loop for efficient data sharing. We deployed and performed an experimental evaluation of \projectName for medical research data. In particular: (i) we compared \projectName data loading and query execution times with eager and lazy ETL, and (ii) we evaluated the efficiency of \projectName in terms of the amount of data replication and bandwidth required in data sharing. The results obtained indicate that \projectName performs better than or equal to both eager and lazy ETL approaches. We further observed that \projectName data sharing feature avoids data replication and repeated ETL efforts.

\paragraph*{\textbf{Paper organization:}}~
The rest of this paper is structured as follows: Section~\ref{sec:arch} presents the solution architecture of \projectName. Section~\ref{sec:impl} describes the implementation details of the \projectName prototype. Section~\ref{sec:eval} presents the experimental evaluation that we conducted and the results obtained. Section~\ref{sec:related} discusses the related work on data integration, data sharing platforms, and ETL approaches for scientific research. Finally, Section~\ref{sec:conclusion} concludes with a summary of the current status and future research directions.

\section{\projectName: An On-Demand Big Data Integration Platform}
\label{sec:arch}
The \projectName platform is instantiated for each organization. Users from the organization can access, integrate, and load data into the integrated data repository of the corresponding \projectName instance. Furthermore, they can share datasets stored in the integrated data repository with other users from the same or different organizations. Section~\ref{subsec:overview} presents the \projectName hybrid ETL approach and the underlying architecture. Section~\ref{subsec:hil} explains how \projectName incorporates human knowledge in the ETL process to selectively and incrementally integrate and load subsets of data and metadata. Section~\ref{subsec:sharing} details \projectName efficient data sharing mechanism beyond organization boundaries to minimize data replication and repeated ETL efforts. 

\subsection{Hybrid ETL Process}
\label{subsec:overview}

\begin{figure}[ht]
    \begin{center}
        \resizebox{\columnwidth}{!}{
            \includegraphics[width=\textwidth]{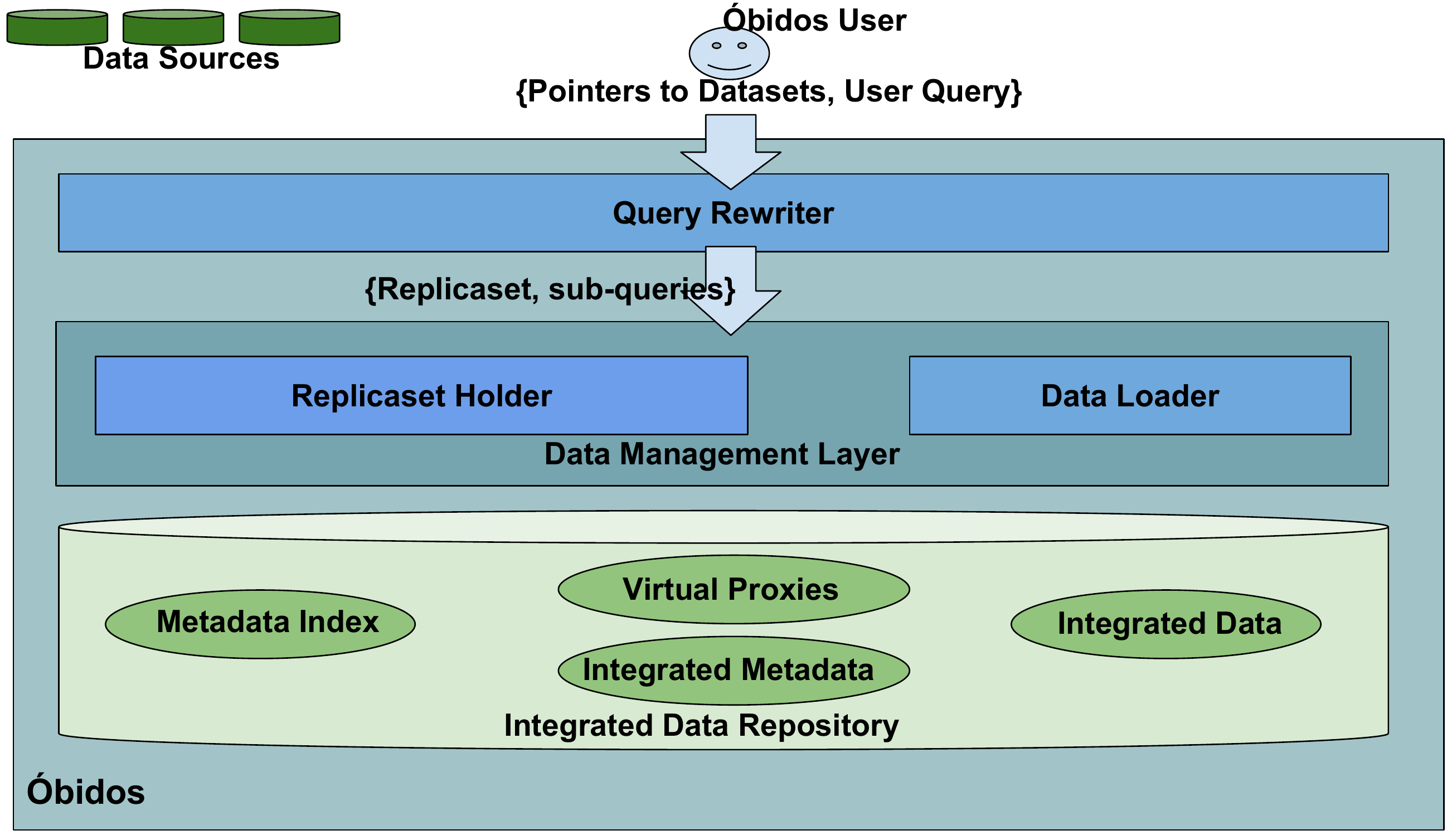}
        }
    \end{center}
    \caption{\projectName Architecture}
    \label{fig:arch}
\end{figure}

\paragraph*{\textbf{\projectName Architecture:}}~Figure~\ref{fig:arch} depicts the architecture of an \projectName instance. From bottom to top, \projectName consists of i) a scalable \textbf{Integrated Data Repository}, ii) a \textbf{Data Management Layer} with constructs for fast data integration and loading, and iii) a \textbf{Query Rewriter} with constructs for efficient and unified access to the data in the integrated data repository and the data sources.

The \textbf{Integrated Data Repository} incrementally stores the data and metadata integrated and loaded by users. It consists of i) structured and unstructured data (including binary data) as \textbf{integrated data} and ii) the corresponding metadata as \textbf{integrated metadata}. Furthermore, the metadata stored in the integrated data repository needs to be indexed for efficient query execution over the binary data. We call this index of the integrated data repository, the \textbf{Metadata Index}. The Metadata Index functions as an internal index that is built over the integrated data and metadata in the \projectName instance. \projectName further stores the incomplete metadata entries, the metadata that is being loaded as \textbf{virtual proxies}. The virtual proxies are stored as \textit{future} or a placeholder for the complete metadata in the integrated data repository. The virtual proxies will be replaced by the complete metadata once the entire metadata is loaded. 

The \textbf{Data Management Layer} consists of data structures to manage the data in the integrated data repository and components to access, integrate, and load from the data sources. It stores its data structures in memory in a cluster of machines, aiming to offer fast access to the integrated data while not compromising fault-tolerance. A \textit{virtual replica} is a pointer to a dataset from a distinct data source. A \textit{replicaset} is an \projectName data structure that is composed of several virtual replicas. Thus, each replicaset points to the distributed and diverse datasets relevant to a scientific research study. Furthermore, the replicasets are identified by timestamps. Therefore, the integrated data repository can be periodically updated with the changes or updates to the datasets in the data sources pointed by the replicasets.

The \textbf{Replicaset Holder} is the core module of the Data Management Layer. It identifies each replicaset by a globally unique identifier known as replicasetID. It stores the replicasets in memory in a data structure that maps each item of integrated and loaded metadata into the corresponding replicasets. Thus, it indicates which of the datasets have already been loaded into the integrated data repository, either as integrated metadata and data or as virtual proxies. Moreover, it enables sharing the replicasets among users freely to make the datasets relevant to the scientific research available to other participants. Therefore, it serves as a component that prevents repetitive attempts to access, integrate, and load the same datasets. The \textbf{Data Loader} selectively loads metadata and data from the data sources. The location and the access mechanisms to the data sources are provided by the user and are stored in memory by the Data Loader. 

Finally, the \textbf{Query Rewriter} enables uniform access to data sources as well as to the integrated data repository. It accepts as input a user query and pointers to the relevant datasets. Then, it converts the pointers to the datasets into replicasets. It also translates user queries into sub-queries that access either the data sources or the integrated repository. If the data required to answer the user query is not present in the integrated data repository, it invokes the Data Loader to integrate and load the datasets to answer the user query as well as the virtual proxies corresponding to the replicaset.

\paragraph*{\textbf{\projectName Incremental Data Integration and Loading:}}~\projectName accesses data and metadata from the data sources, and incrementally integrates and loads the results of the user queries into an integrated data repository. The integrated data repository persists previous query answers as well as the data and metadata integrated and loaded for answering previous queries. Therefore, queries can be regarded as virtual datasets that can be re-accessed or shared (akin to the materialized view in traditional RDBMS). 

\projectName enables to incrementally integrate and load metadata to mitigate the challenges in loading the metadata entirely or eagerly.
When incrementally loading the metadata, \projectName replaces the counterparts of metadata that has not been loaded yet with a placeholder. We call such partially loaded metadata, a virtual proxy to the actual metadata. The use of virtual proxies minimizes the volume of metadata integrated and loaded. \projectName stores the virtual proxies in the integrated data repository in addition to the integrated data and the corresponding metadata. If only a fraction of metadata is relevant for a search query, it is sufficient to load only that fraction. Therefore, \projectName selectively loads metadata as virtual proxies. The virtual proxies are later replaced by the complete metadata as the metadata is accessed and integrated. Thus, virtual proxies refer to the metadata of a dataset larger than that is integrated and loaded to the integrated data repository. 

Often a virtual replica may be present in the Replicaset Holder, without having the exact data for the user query. This usually means, previously at least one different user query has been executed on the same virtual replicas. Therefore, while the virtual proxies of the replicaset are present, the exact data for the user query may not be present in the integrated data repository. With time, as more and more data are selectively integrated and loaded, the integrated data repository will contain the necessary data for the subsequent scientific research queries.

\subsection{Human-in-the-Loop ETL Process}
\label{subsec:hil}

\projectName supports a human-in-the-loop ETL process. By `human-in-the-loop', here we mean to incorporate the human knowledge that corresponds to the user-defined replicasets and queries to selectively access and integrate data from the data sources and incrementally loading the integrated data repository. A user identifies certain datasets as relevant to her scientific research, and these datasets are the ones against which the user query will be executed. She defines a replicaset by including pointers to these datasets as virtual replicas. The replicaset and a specific user query define the data to be integrated and loaded by each selective data integration and loading process. This avoids the need to exhaustively look for the desired data across data sources.  

The \projectName selective load process is initiated every time a user issues a query. First, \projectName iteratively checks for the existence of the data necessary to answer the query in the instance. It queries the Replicaset Holder for each of the virtual replicas and then executes the user query on the integrated data repository. If the data is not available in the instance, it is integrated and loaded from the data sources. The results of the user queries are persistently stored into the integrated data repository. Furthermore, rather than just querying and loading only the answers of the user query, \projectName selectively loads the metadata pointed by the replicaset. This ensures that the integrated data repository can be incrementally loaded with data, rather than merely storing discrete, incoherent, or independent sets of data.

\begin{figure}[ht]
    \begin{center}
        \resizebox{\columnwidth}{!}{
            \includegraphics[width=\textwidth]{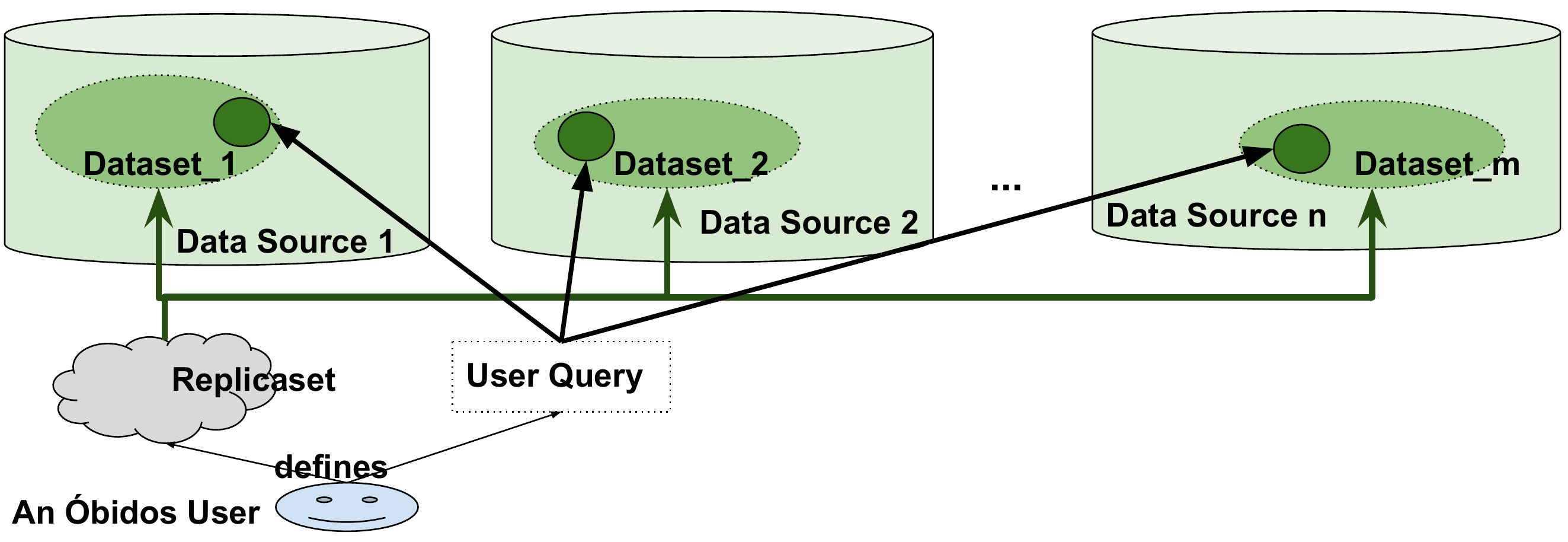}
        }
    \end{center}
    \caption{Narrowing down the search space with user-defined replicasets}
    \label{fig:replicaset}
\end{figure}

Figure~\ref{fig:replicaset} shows an \projectName user defining a replicaset along with a user query to be executed on multiple data sources. The replicaset narrows down the search space from the entire data sources to specific datasets to answer the user query. She ensures with the knowledge of the data sources, the data required to answer her user query is part of the datasets pointed by her replicaset.

The data integrated and loaded into the integrated data repository of an \projectName instance should be available to be accessed later for scientific research. For example, when a user receives a replicaset from another user from the same or another organization, she may access her organization's instance to check for already loaded data. 

Algorithm~\ref{alg:hybrid} illustrates how a user initiates the selective and incremental data integration and loading process of \projectName. The font color in the algorithm represents the nature of the execution. Red represents conditions, blue represents data manipulations, and green represents general computing executions.

\begin{algorithm}[ht]
        \fontsize{9}{9}\selectfont
    \caption{\projectName Human-in-the-Loop Incremental ETL}
    \label{alg:hybrid}
    \begin{algorithmic}[1]

        \Procedure{selectiveLoad}{\textit{replicaset, userQuery}} 
          \State \colorbox{blue!10}{\textit{toLoad} $\gets$ \textit{replicaset}}         
    \ForAll {\colorbox{red!10}{\textit{(virtualReplica} \textbf{in} \textit{replicaset)}}}
    
          \State \colorbox{green!10}{\textit{wasLoadedBefore} $\gets$ \textit{replicasetHolder.}\textbf{get}\textit{(virtualReplica)}} 
           \If { \colorbox{red!10}{(\textit{!(wasLoadedBefore)}}}
                   \State \colorbox{blue!10}{\textit{loadData(virtualReplica, userQuery)}}
\State \colorbox{blue!10}{\textit{replicasetHolder.} \textbf{put} \textit{(virtualReplica)}}                    
          \State \colorbox{blue!10}{\textit{toLoad}.\textbf{delete(}\textit{virtualReplica})}                          
             
                 \EndIf   \EndFor

    \If { \colorbox{red!10}{\textit{((toLoad.}\textbf{size} $>$ 0) \textbf{AND}}
    \par\hspace{30pt}{\colorbox{red!10}{ \textit{(integratedDataRepository.query(userQuery) == NULL))}}}}
    
        \ForAll {\colorbox{red!10}{\textit{(virtualReplica} \textbf{in} \textit{toLoad)}}}
                   \State \colorbox{blue!10}{\textit{loadData(virtualReplica, userQuery)}}
\EndFor
                           \EndIf  
        \EndProcedure
    \end{algorithmic}

\end{algorithm}

The algorithm starts by initializing a temporary variable \textit{toLoad}, as a set, with the copy of the replicaset (line 2). \textit{toLoad} tracks the virtual replicas belonging to the replicaset that have not yet been loaded from the data sources. Then, the algorithm proceeds to check the existence of the data pointed by each virtual replica in the instance (line 3). First, it queries the Replicaset Holder to check whether datasets pointed by the virtual replica have already been loaded by a previous query (line 4). If no dataset has yet been loaded for the virtual replica (line 5), the data relevant for the virtual replica and the user query is loaded from the data sources incrementally, invoking the \textit{loadData} procedure (line 6). The Replicaset Holder matches the replicasets to the respective data and metadata integrated and loaded in the integrated data repository, by the selective load process. Therefore, in line 7, the virtual replica is added to the Replicaset Holder. Now since the dataset pointed by the virtual replica has already been loaded, the virtual replica is removed from \textit{toLoad} (line 8). 

The first loop (lines 3 - 10) checks whether the data, metadata, or virtual proxies relevant for one or more of the virtual replicas exist in the integrated data repository. It loads the data only when neither corresponding data and metadata nor virtual proxies are found for a given virtual replica. Therefore, a non-empty set of \textit{toLoad} at the end of the loop indicates that at least a few virtual replicas were not loaded during this iteration. In that case, the algorithm proceeds to check whether the data and metadata necessary to answer the current user query are completely available in the integrated data repository (line 11). The user query will return a NULL if the complete metadata and data necessary to answer the query are not present in the integrated data repository. Consequently, the \textit{loadData} procedure is executed for all the virtual replicas in the \textit{toLoad} set (lines 12 - 14).

\paragraph*{\textbf{The loadData Procedure:}}~The \textit{loadData} procedure is the core of the \projectName human-in-the-loop incremental ETL approach. It accepts a replicaset and a user query as input arguments. First, the data sources are accessed, and the datasets identified by the replicasets are selectively loaded as virtual proxies, without loading the entire metadata. Then, the user query is executed against the data sources. The relevant metadata representing the results of a user query is integrated and loaded to the integrated data repository. If the user query also indicates access to the binary data, the respective binary data (usually a subset of data corresponding to the metadata already loaded by the query) is also loaded to the integrated data repository. The \textit{loadData} procedure selectively loads the metadata corresponding to the replicaset as virtual proxies. If previously a different user query was issued with the same virtual replica, the virtual proxies corresponding to the virtual replica would be present while the exact data and metadata to answer the current user query would remain absent in the integrated data repository.

\subsection{Data Sharing Process}
\label{subsec:sharing}

An \projectName instance is deployed in each organization. Each \projectName instance is used by: i) users from the organization, and ii) users from other organizations and external users who have limited access to the \projectName instance. Users can share the datasets among them by sharing the replicasets or their respective replicasetIDs. Therefore, there is no need to replicate the actual data of the data sources nor the integrated data repository of an \projectName instance.

Replicasets are small in size. However, they grow with the number of data sources and diversity of data. ReplicasetID is smaller in size compared to the replicaset and is of a fixed size. Therefore, they are shared by default. A user outside the organization can access the data already loaded in an \projectName instance using the replicasetID. Moreover, users can share the replicasets with other organizations, without letting them access to the data in their integrated data repository. The organizations can then integrate and load the datasets pointed by the replicaset, from the data sources. The relevant datasets pointed by the received replicaset can later be integrated and loaded by the remote users to their own \projectName instance. 

\begin{figure}[ht]
    \begin{center}
        \resizebox{\columnwidth}{!}{
            \includegraphics[width=\textwidth]{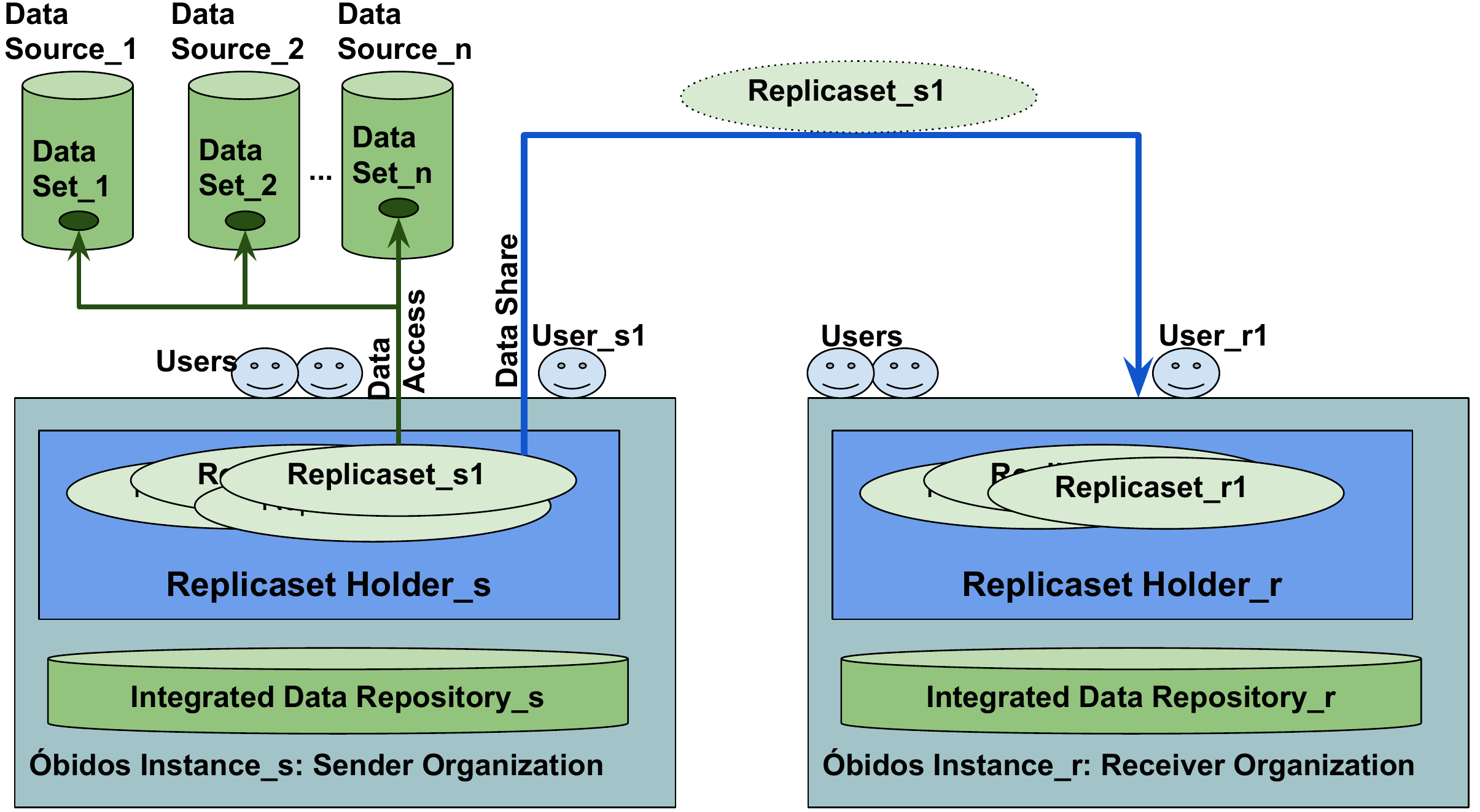}
        }
    \end{center}
    \caption{Data Sharing with \projectName}
    \label{fig:share}
\end{figure}

Figure~\ref{fig:share} illustrates the process of data sharing between users \textit{User\_s1} and \textit{User\_r1} from two different organizations (called sender and receiver). The sender organization and the receiver organization can also represent the same organization if both users belong to the same organization. Datasets can be shared by as a replicaset or the respective replicasetID.

Algorithm~\ref{alg:start} describes the data sharing procedure executed by the \projectName instance of the receiver organization. It takes as input: a replicaset (or its replicasetID) received from another user, the identification of users that created/sent and received the replicaset, and an object indicating whether the datasets need to be accessed directly from the sender instance (defined as the \textit{accessSender}) (line 1). The \textit{accessSender} consists of a boolean flag, along with the relevant access mechanisms such as the access key to the integrated data repository of the sender instance. It indicates whether the integrated data repository of the sender instance should be accessed directly by the receiver.

\begin{algorithm}[ht]
        \fontsize{9}{9}\selectfont
    \caption{Data Sharing via a Replicaset}
    \label{alg:start}
    \begin{algorithmic}[1]

        \Procedure{shareReplicaset}{\textit{replicaset, sender, receiver, accessSender}}
               
           \If { \colorbox{red!10}{\textit{(replicaset.isURI())} }
                   \State \colorbox{green!10}{\textit{replicaset} $\gets$ \textit{sender.get(replicaset)}}}
                 \EndIf
    
               \If { \colorbox{red!10}{\textit{(accessSender)} }
                   \State \colorbox{blue!10}{\textit{sender.access(replicaset)}}}
                \Else 
                       \State \colorbox{blue!10}{ \textit{receiver.selectiveLoad(replicaset, NULL)}}            
                 \EndIf
        \EndProcedure
    \end{algorithmic}
\end{algorithm}

If a replicasetID is received, the replicaset is retrieved from the sender instance first (lines 2 - 4). Since the replicaset was initially created by a user of the sender organization, the datasets or the virtual proxies pointed by the replicaset would be present in the sender organization. Therefore, if the \textit{accessSender} is set to a non-null value (line 5), the datasets pointed by the replicaset are accessed directly from the sender instance, by the receiver organization (line 6). Otherwise, the \textit{shareReplicaset} procedure selectively loads the datasets pointed by the replicaset into the receiver instance, from the data sources (line 8). As there is no user query defined in a shared replicaset, the \textit{selectiveLoad} procedure is invoked with a null value in place of the user query.

\section{Implementation}
\label{sec:impl}
We built each of the \projectName processes, including data cleaning, loading, and sharing, as a service. Thus, \projectName builds the hybrid ETL with data sharing as a chain of data services with associated data structures.

\subsection{Data Structures}
\begin{figure}[ht]
    \begin{center}
        \resizebox{\columnwidth}{!}{
            \includegraphics[width=\textwidth]{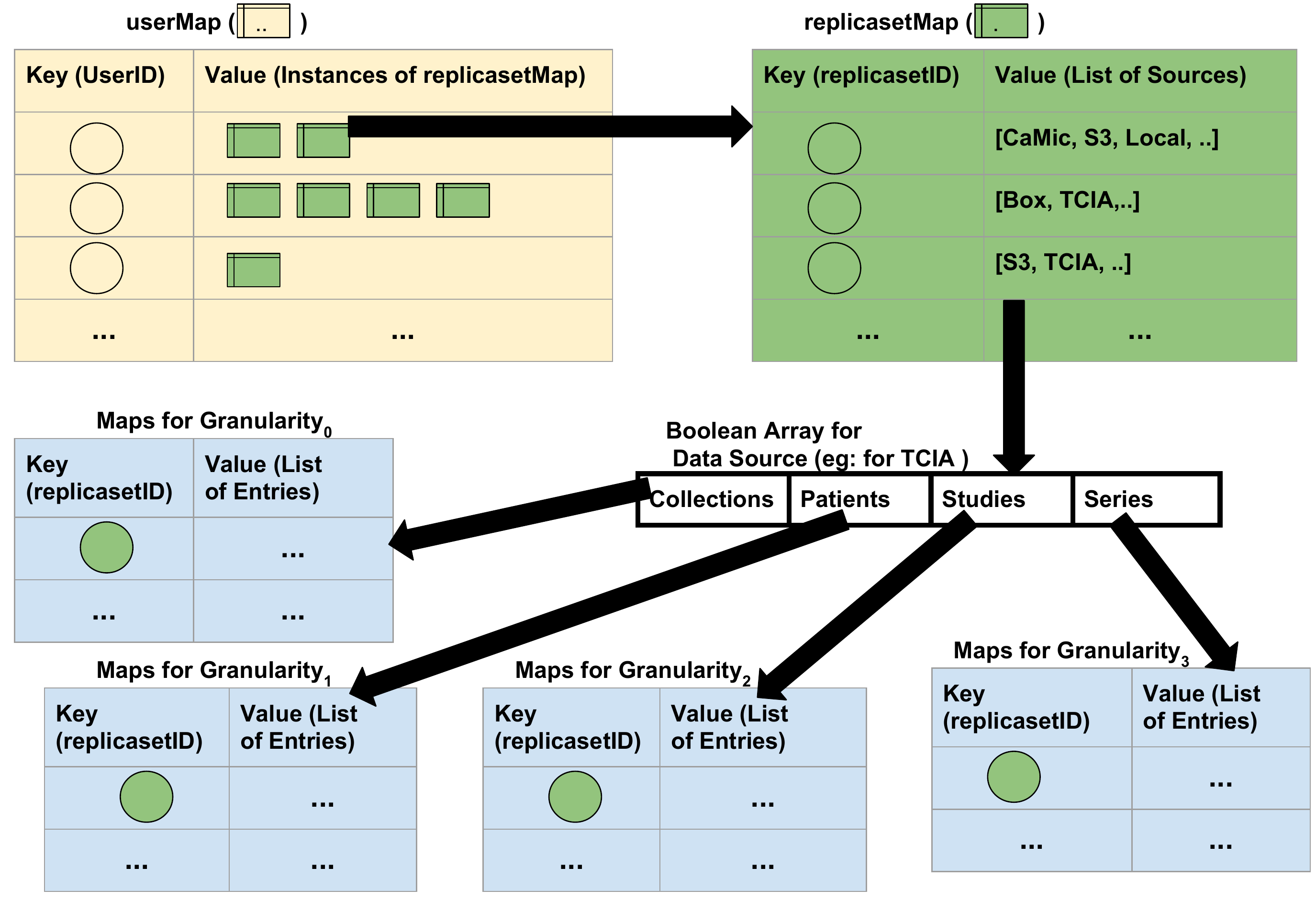}
        }
    \end{center}
    \caption{Data Structures of the Replicaset Holder}
    \label{fig:datarep}
\end{figure}

Figure~\ref{fig:datarep} illustrates the data representation of \projectName. The maps that represent each granularity resolve to store the identifier of the metadata or a virtual proxy. \projectName presents the replicasets internally in a minimal tree-like data structure. To offer efficient search and indexing capabilities, virtual proxies and metadata are built into a hierarchical map structure. The Replicaset Holder consists of a few instances of the \textit{multi-map} data structure, where a set of items is stored as the value in the map, against a given key. As each user composes several replicasets, the \textbf{userMap} stores a list of replicasets against the identification of the users (userID) that created them. Each entry in the list of values of the \textbf{userMap} represents a replicaset of a user. 

The specific contents of the replicasets are stored in a \textbf{replicasetMap}, including the virtual replicas belonging to each replicaset, and whether the replicasets have already been integrated and loaded to the integrated data repository. Replicasets include pointers to datasets from various data sources as virtual replicas. Therefore, each \textit{replicasetMap} indicates the relevant data sources for each of the replicaset. It employs the replicasetID as its key and the list of data source names contributing data to a replicaset as the value. 

Each data source belonging to a replicaset is internally represented by $n$ maps. In a hierarchical data storage format such as DICOM, each of the $n$ maps represents one of the granularity levels in the data source. Such a format facilitates seamless integration of virtual proxies into the metadata. A boolean array $A_b$ of length $n$ is used to represent the replicaset in a bit-map like manner. Each element $A_b[i]$ of the array represents the existence of a non-null entry in the $i^{th}$ map of granularity. Thus, the boolean flags in $A_b$ indicate the existence (or lack thereof) of the dataset in a particular granularity. If an entire level of granularity is included in the replicaset by the user, the relevant flag is set to true. 

Figure~\ref{fig:datarep} represents an illustrative use case for a hierarchical data storage. It considers cancer images of DICOM format stored in data repositories such as TCIA, S3 buckets, directories in Box.com, and a local folder/file hierarchy. For these cancer images of DICOM format, $n$ = 4. Thus, a map represents each of its 4 granularity levels - collections, patients, studies, and series, with an array of length 4 pointing to each of the 4 maps. The hierarchical data representation enables incremental loading and virtual proxies. Thus \projectName offers a fast and indexed data structure to access the metadata and data loaded into the integrated data repository.

\subsection{Service-based APIs}
\projectName is built as a service-based hybrid ETL with RESTful service interfaces. It offers a web services interface for its hybrid ETL and data sharing. The \projectName hybrid ETL is designed as CRUD (Create, Retrieve, Update, and Delete) functions on replicasets. These functions are exposed as RESTful services, POST, GET, PUT, and DELETE. 

\projectName offers a \textit{data sharing} API to share scientific research datasets, by sharing the replicasets. Replicasets can also be shared outside \projectName, through other communication media such as email. The data sharing method is typically one-to-one, meaning that a user shares data with another user in the same or different organization. However, it can also be listed for the public to be freely accessed.

The user accesses, queries, integrates, and loads the relevant data from the data sources by invoking the \textit{\textbf{create} replicaset} procedure. This procedure creates a replicaset and initiates the selective data integration and loading process. When \textit{\textbf{retrieve} replicaset} is invoked, the data corresponding to the given replicaset is retrieved from the integrated data repository. Furthermore, \projectName checks for updates from the data sources pointed by the replicaset, if the data corresponding to the replicaset has already been integrated and loaded. Metadata of the replicaset is compared against that of the data sources for any corruption or local changes. The user deletes existing replicasets by invoking the \textit{\textbf{delete} replicaset}. The Replicaset Holder is updated immediately to avoid loading updates to the deleted replicasets. The user updates an existing replicaset to increase, decrease, or alter its scope, by invoking the \textit{\textbf{update} replicaset}. This may, in turn, invoke parts of create and delete processes, as new data may be loaded while existing parts of data may be removed.

The Replicaset Holder associates data with a user. It thus virtually associates each dataset to a user, through its data structures such as the userMap. While each user has her own virtually isolated space in memory, the integrated data repository consists of a data storage shared among all the users of the organization. Hence, before deleting a data entry from the integrated data repository, the data should be confirmed to be an `orphan' with no replicasets referring to them from any of the users. Deleting data from the integrated data repository is designed to be initiated by a background task, rather than its regular users. When the storage is abundantly available in a cluster, \projectName advocates keeping orphan data in the integrated data repository rather than immediately initiating the cleanup process, and repeating it too frequently.

\subsection{\projectName Software Components}
We leverage several software frameworks in building the \projectName prototype. Apache Hadoop Distributed File System (HDFS)~\cite{white2012hadoop} is used as the core of the integrated data repository, due to its scalability and support for storing unstructured and semi-structured, binary and textual data. The execution is performed on a cluster of Infinispan~\cite{marchioni2012infinispan} in-memory data grid. Consequently, data structures of the Data Management Layer are stored in an Infinispan cluster. The metadata of the binary data in HDFS is stored in tables hosted in Apache Hive~\cite{thusoo2009hive} metastore based on HDFS. The Hive tables consisting of the metadata are indexed with the Metadata Index for users to query and locate the data from the integrated data repository efficiently.

Apache Drill enables SQL queries on structured, semi-structured, and unstructured data. Therefore, the Query Rewriter unifies and accesses the storages seamlessly by leveraging Apache Drill~\cite{hausenblas2013apache}. Thus, \projectName supports SQL queries on unstructured data stored in HDFS, by leveraging the Metadata Index stored in Hive. This approach allows efficient queries to the data, partially or wholly loaded into the integrated data repository. Thus, \projectName provides unified and scalable access to the data in the integrated data repository and the data sources. 

Oracle Java 1.8 is used as the programming language in developing \projectName. Apache Velocity 1.7~\cite{gradecki2003mastering} is leveraged to generate the application templates of the \projectName web interface. Hadoop 2.7.2 stores the integrated data along with its corresponding metadata and virtual proxies, while the Metadata Index is stored in Hive 1.2.0. Hive-jdbc package writes the Metadata Index into the Hive metastore through its JDBC bindings to Hive. SparkJava 2.5~\cite{sparkjava} compact Java-based web framework is leveraged to expose the \projectName APIs as RESTful services. The APIs are managed and made available to the relevant users through API gateways. API Umbrella is deployed as the default API gateway. \projectName incorporates authorization to its shared data from the integrated data repository through the use of API keys, leveraging the API gateway. Thus, one can only access the data shared with them, and only with the API key that belongs to them. 

Embedded Tomcat 7.0.34 is used to deploy \projectName as a web application. Infinispan 8.2.2 is used as the In-Memory Data Grid where its distributed streams support distributed execution of \projectName processes across the \projectName clustered deployment. The data structures of the Data Management Layer are represented by instances of the Infinispan Cache class, which is a Java implementation of distributed HashMap. Drill 1.7.0 is exploited for the SQL queries on the integrated data repository, with drill-jdbc offering JDBC API to interconnect with Drill from the Query Rewriter.

\section{Evaluation}
\label{sec:eval}
\projectName has been benchmarked against implementations of eager ETL and lazy ETL approaches, using microbenchmarks derived from medical research queries on cancer imaging and clinical data. 

\begin{figure}[ht]
    \begin{center}
        \resizebox{\columnwidth}{!}{
            \includegraphics[width=\textwidth]{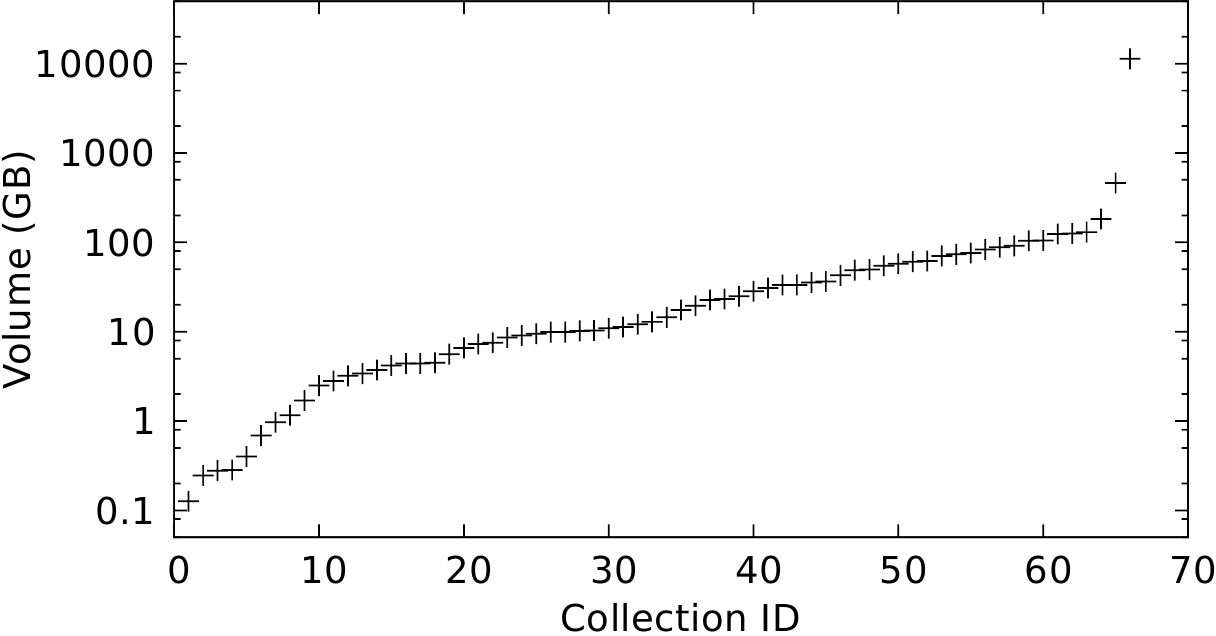}
        }
    \end{center}
    \caption{Evaluated DICOM Imaging Collections (Sorted by Total Volume)}
    \label{fig:colsize}
\end{figure}

\paragraph*{\textbf{Evaluation Environment and Benchmark Analysis:}}~
An \projectName prototype, implemented as described above, has been deployed to integrate medical data from various heterogeneous data sources including The Cancer Imaging Archive (TCIA)~\cite{clark2013cancer}, DICOM imaging data hosted in Amazon S3 buckets, medical images accessed through caMicroscope~\cite{camicroscope}, clinical and imaging data hosted in local data sources including relational and NoSQL databases, and file system with files and directories along with CSV files as metadata. The core data used in the evaluations are DICOM images. They are stored as collections of various volume as shown in Figure~\ref{fig:colsize}. The data consists of large-scale binary images (in the scale of a few thousand GB, up to 10 000 GB) along with a smaller scale textual metadata (in the range of MBs).

\begin{figure}[ht]
    \begin{center}
        \resizebox{\columnwidth}{!}{
            \includegraphics[width=\textwidth]{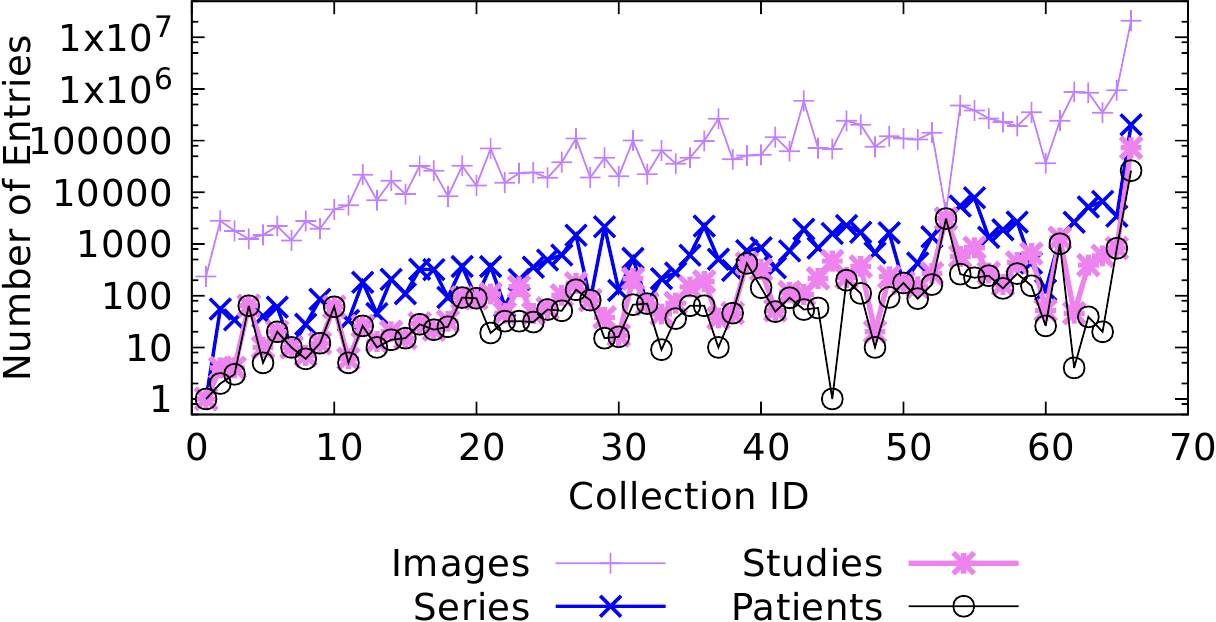}
        }
    \end{center}
    \caption{Various Entries in Evaluated Collections (Sorted by Total Volume)}
    \label{fig:entries}
\end{figure}

Figure~\ref{fig:entries} illustrates the number of patients, studies, series, and images in each of the collection. Collections are sorted according to their total volume. Each collection consists of multiple patients; each patient has one or more studies; each study has one or more series; and each series has multiple images. We defined replicasets at these different levels of granularity. The varying pattern of Figure~\ref{fig:entries}, when compared against that of Figure~\ref{fig:colsize} shows that the total volume of a collection does not necessarily reflect the number of entries in it.

\subsection{Performance of Integrating and Loading Data}
\projectName was benchmarked for its performance and efficiency in integrating and loading the data. \projectName integrates and loads data from the scientific research data sources spanning across the globe. Therefore, the performance of loading the data will be influenced by the bandwidth. To avoid this influence, first, we replicated the data sources such as TCIA to data sources hosted on the local servers. 

\begin{figure}[ht]
    \begin{center}
        \resizebox{0.9\columnwidth}{!}{
            \includegraphics[width=0.9\textwidth]{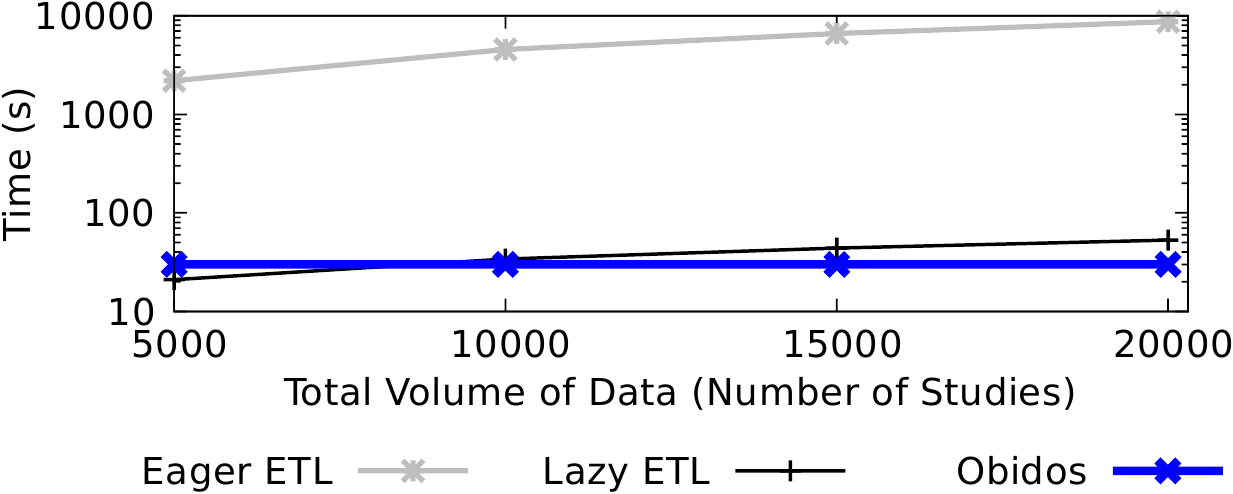}
        }
    \end{center}
    \caption{Data load time: change in total volume of data sources (Same user query and same replicaset)}
    \label{fig:totalsize}
\end{figure}

We integrated and loaded data from different total volumes of data sources for the same replicasets of the user. We measured the volume of the data sources by the total number of studies in them. Figure~\ref{fig:totalsize} shows the data load time of \projectName against that of lazy ETL and eager ETL approaches. Since \projectName selectively loads the metadata of only the data corresponding to the replicaset, the loading time remained constant independent of the increasing total volume of data in the data sources. However, since lazy ETL and eager ETL approaches query the entire data sources, the increase of volume leads to a larger time to integrate and load them. Eager ETL always took more time as it has to integrate and load the entire metadata and data. Since lazy ETL loads only the metadata eagerly, it loads faster than eager ETL.

Furthermore, for smaller volumes of data, eagerly loading the entire metadata can be faster than the selective loading by \projectName, as \projectName executes the query on the data source and loads the virtual proxies, creating and updating the constructs such as the Metadata Index and the Replicaset Holder. Therefore, \projectName took longer for the data integration and loading compared to the lazy ETL for smaller volumes of data. However, as the total volume of data grows, the data loaded by \projectName remained the lowest, compared to both eager ETL and lazy ETL. Moreover, for repeating user queries, both eager ETL and \projectName outperformed the lazy ETL due to the availability of the integrated data repository in both eager ETL and \projectName, and the storing of query answers in \projectName.

\begin{figure}[ht]
    \begin{center}
        \resizebox{0.9\columnwidth}{!}{
            \includegraphics[width=0.9\textwidth]{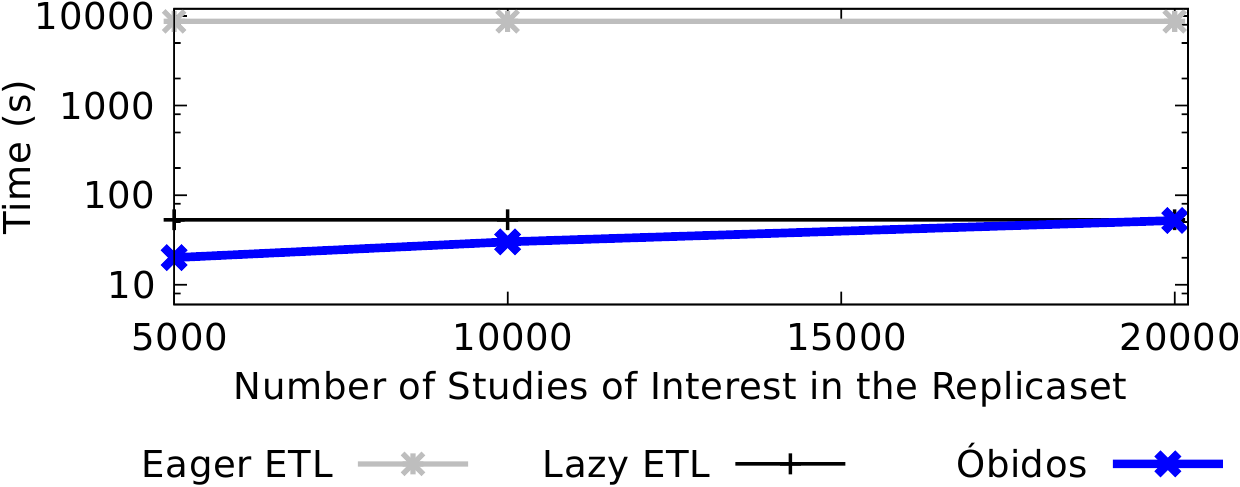}
        }
    \end{center}
    \caption{Data load time: varying number of studies of interest in the replicaset (same user query and constant total data volume)}
    \label{fig:interest}
\end{figure}

The experiment was repeated for a constant total volume of data sources while increasing the number of studies of interest in the replicaset. Figure~\ref{fig:interest} shows the time taken by \projectName, lazy ETL, and eager ETL to integrate and load the data from the data sources. Since the total volume remained constant, the lazy ETL and eager ETL had the same data integration and loading time, as they remain oblivious to the change in the number of studies of interest. However, the performance of \projectName depends heavily on how the replicasets are defined. Therefore, with the growth of the replicaset, the loading time of \projectName increased. Eventually, the data integration and loading time of \projectName converged with the time taken by the lazy ETL approach, as the replicaset was defined to cover all the studies in the data sources (thus, making it eagerly loading the metadata).

\begin{figure}[ht]
    \begin{center}
        \resizebox{0.9\columnwidth}{!}{
            \includegraphics[width=0.9\textwidth]{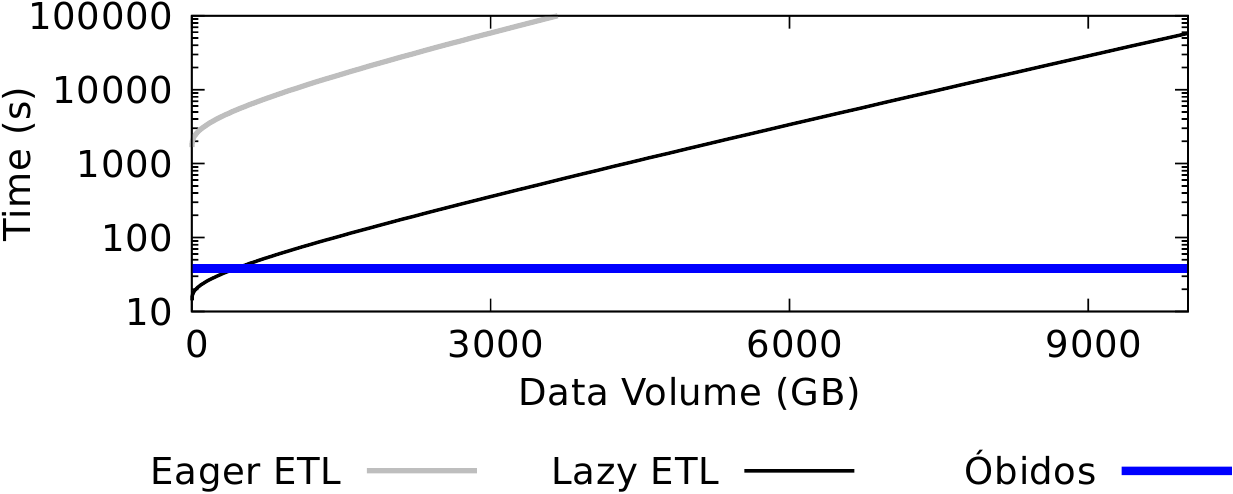}
        }
    \end{center}
    \caption{Load time from the remote data sources}
    \label{fig:download}
\end{figure}

Finally, datasets were integrated and loaded directly from the remote data sources (such as TCIA and S3 buckets) through their web service APIs, to evaluate the effects of data downloading and bandwidth consumption associated with it. We changed the total volume of data in the data sources by adding more data to the data sources while keeping the replicaset unchanged. Figure~\ref{fig:download} shows the time taken for \projectName, lazy ETL, and eager ETL. Eager ETL performed poor as binary data had to be downloaded over the network. Lazy ETL too performed slowly for large volumes as it must eagerly load the metadata (which itself grows with scale) over the network. 

As with the case of Figure~\ref{fig:totalsize}, Figure~\ref{fig:download} too illustrates a fixed time for \projectName data integration and loading. As the data was integrated and loaded over the Internet from the data sources, the time taken grew linearly for eager ETL and lazy ETL. However, lazy ETL consumed much lower time compared to the eager ETL. As only the datasets corresponding to the replicaset are accessed, integrated, and loaded, \projectName uses bandwidth conservatively, loading no irrelevant data or metadata. Regardless of the growth of the increasing total volume of data in the data sources, \projectName integrated and loaded the data at the same time as the replicaset and the user query remained the same. Therefore, the human-in-the-loop contributed positively to the integration and loading performance of \projectName by narrowing down the search space from the data sources.

\subsection{Performance of Querying the Integrated Data Repository}
\begin{figure}[ht]
    \begin{center}
        \resizebox{\columnwidth}{!}{
            \includegraphics[width=\textwidth]{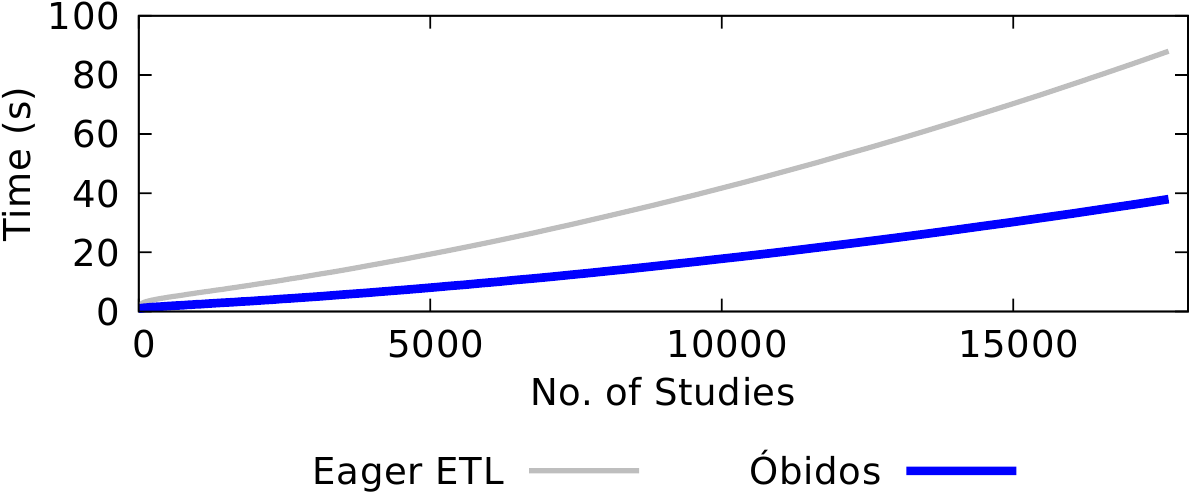}
        }
    \end{center}
    \caption{Query completion time for the integrated data repository}
    \label{fig:client}
\end{figure}

\projectName was then benchmarked for its efficiency in querying the data and integrated data repository against the eager ETL. Query completion time depends on the number of entries in the queried data rather than the size of the entire integrated data repository. Hence, varying amounts of data, measured by the number of studies, were queried. The query completion time of \projectName and eager ETL is depicted in Figure~\ref{fig:client}. \projectName showed a speedup compared to the eager ETL, which can be attributed to the efficient indexing of the integrated data repository with the binary data with Metadata Index and the efficiency of the Data Management Layer in managing the storage and execution. The unstructured data in HDFS was very efficiently queried as in a relational database through the distributed query execution of Drill with its SQL support for NoSQL data sources. 

Typically, lazy ETL approaches do not consist of an integrated data repository. Therefore, we avoid comparing the query performance on the \projectName integrated data repository against the lazy ETL. Eager ETL could outperform \projectName for queries that access data not yet loaded in \projectName, as eager ETL would have constructed an entire data warehouse beforehand. However, with the domain knowledge of the medical data researcher, the relevant datasets are loaded timely, and only those. The time required to construct a complete data warehouse would preclude any benefits of eager loading from being prominent. If data is also not loaded beforehand in eager ETL, it will consume much longer to construct the entire data warehouse before actually starting the processing of the user query. Moreover, loading everything beforehand may be irrelevant, impractical, or even impossible for scientific research studies due to the scale and distribution of the data sources.

Overall, in all the relevant use cases, lazy ETL and \projectName significantly outperformed eager ETL as the need to build a complete data warehouse is avoided in them. As \projectName loads only the relevant subsets of metadata, and does not eagerly load even the metadata, for large volumes \projectName also significantly outperformed lazy ETL in its integration and loading. In addition, the human-in-the-loop selective ETL approach of \projectName satisfies the requirement of the scientific research to have protected access to the sensitive data.

\subsection{Sharing Efficiency of Medical Research Data}

\begin{figure}[!ht]
        \subfloat[With changing number of shared series\label{fig:sharing-a}]{\includegraphics[width=0.7\textwidth]{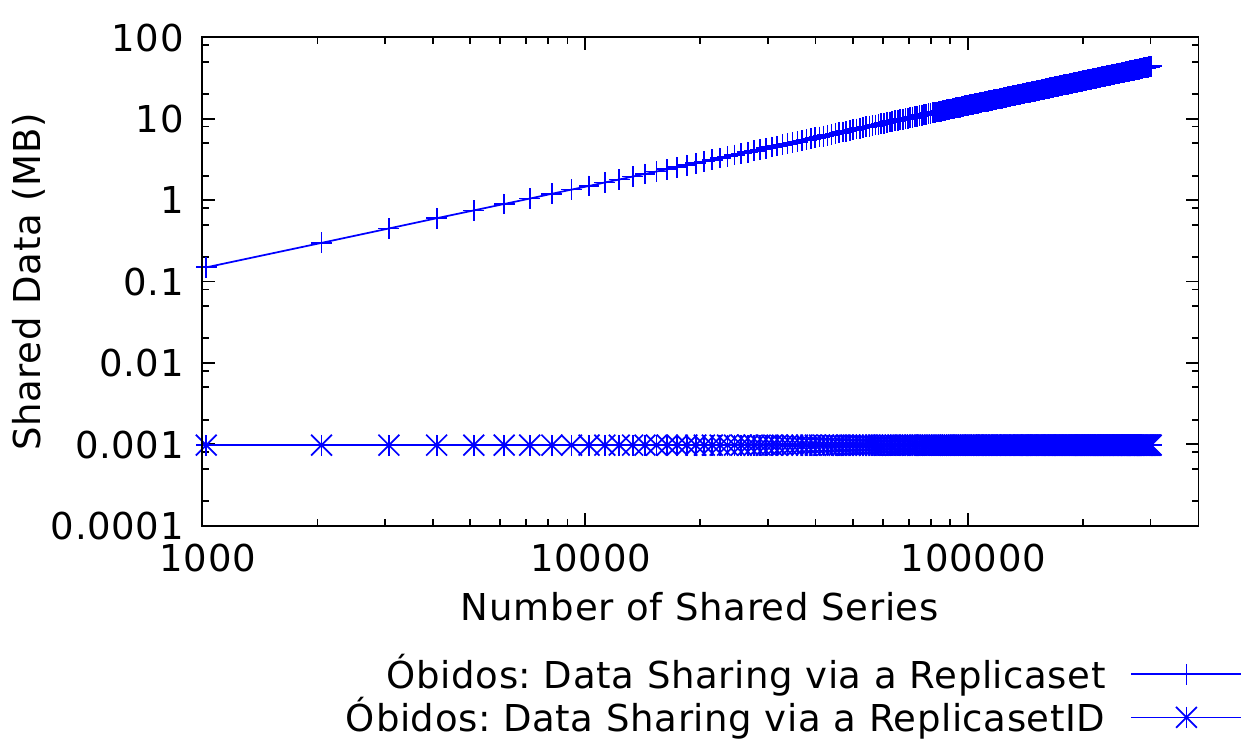}}\\
          \subfloat[With changing volume of shared images\label{fig:sharing-b}]{\includegraphics[width=0.7\textwidth]{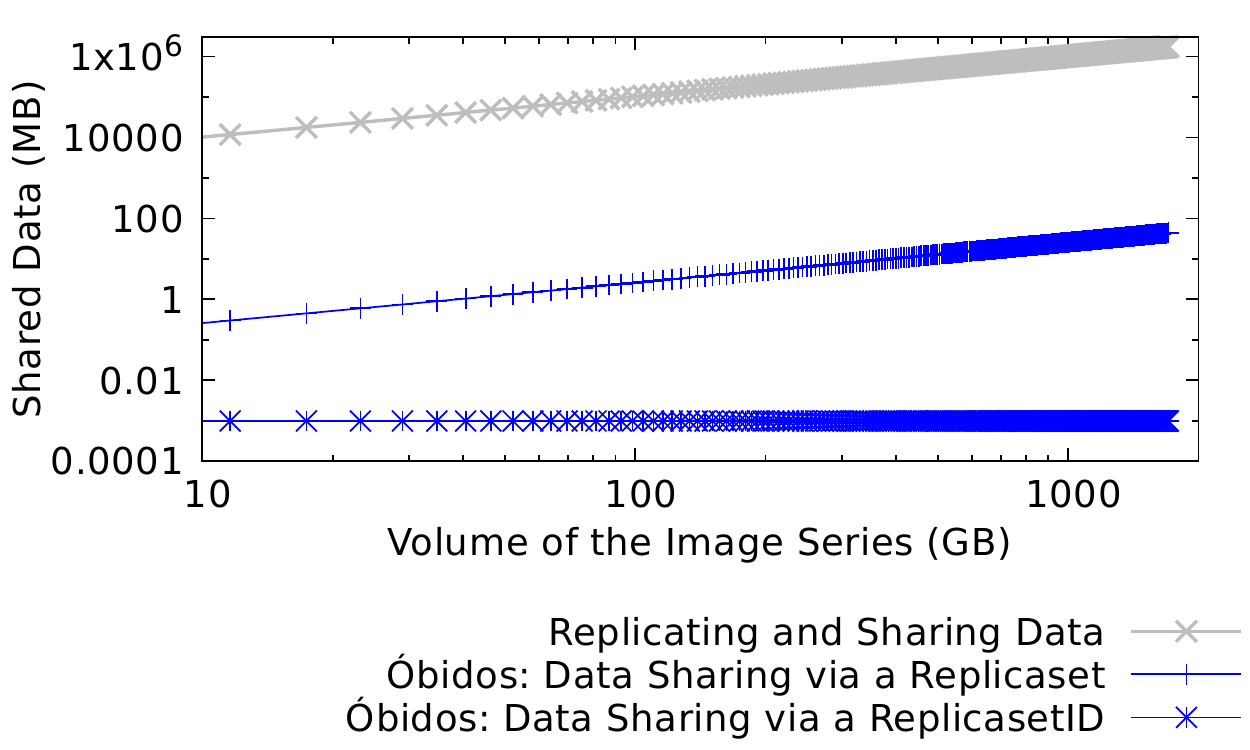}}
        \caption{Volume of data shared in \projectName data sharing use cases vs. in regular binary data sharing}
        \label{fig:sharing}
\end{figure}

Various image series of an average uniform size are shared between users inside an \projectName instance and across multiple instances. Figure~\ref{fig:sharing} benchmarks the data shared in these \projectName data sharing approaches against the typical binary data transfers regarding its bandwidth efficiency. \projectName can share data by sharing either the replicasetID or replicaset. ReplicasetIDs are very small and are fixed in size. Replicasets are minimal in size as pointers to actual data. However, they grow linearly when more data of the same level of granularity is shared. Minimal overhead was added in both cases as compared to sharing actual data. The \projectName data sharing approach also avoids the need for manually sharing the locations of the datasets, which is an alternative bandwidth-efficient approach to sharing the integrated data. As the pointers are shared, no actual data is copied and shared. This enables data sharing with zero redundancy.

The data sharing process of \projectName is designed to have minimal data replication across multiple organizations, avoiding repetitive ETL efforts. Through it is support for sharing datasets through a globally identifiable replicasetID, the data sharing is made efficient with minimal bandwidth overhead. Even sharing the replicaset itself was much bandwidth-efficient than actually replicating and sharing the data. Furthermore, by limiting unauthorized access to the integrated data repository (through authorization mechanisms such as API keys), \projectName avoids accidental sharing of confidential scientific research data. When the receiver does not have access to the integrated data repository of the sender organization, the datasets pointed by the replicaset are integrated and loaded into the receiver organization's integrated data repository.

\section{Related Work}
\label{sec:related}

\paragraph*{\textbf{Service-Based Data Integration:}}~ OGSA-DAI (Open Grid Services Architecture - Data Access and Integration)~\cite{antonioletti2005design} facilitates federation and management of various data sources through its web service interface. The Vienna Cloud Environment (VCE)~\cite{borckholder2013generic} offers service-based data integration of clinical trials and consolidates data from distributed sources. VCE offers data services to query individual data sources and to provide an integrated schema atop the individual datasets. This is similar to \projectName, though \projectName offers a complete hybrid ETL approach and supports sharing of data with minimal data replication.

EUDAT~\cite{lecarpentier2013eudat} is a platform to store, share, and access multidisciplinary scientific research data. EUDAT hosts a service-based data access feature B2FIND~\cite{widmann2016eudat}, and a sharing feature B2SHARE~\cite{ardestani2015b2share}. When researchers access these cross-disciplinary research data sources, they already know which of the repositories they are interested in, or can find them by the search feature. Similar to the motivation of \projectName, loading the entire data from all the sources is irrelevant in EUDAT. Hence, choosing and loading certain sets of data is supported by these service-based data access platforms. \projectName can be leveraged to load related cross-disciplinary data from the eScience data sources such as EUDAT.

\paragraph*{\textbf{Lazy ETL:}}~
Lazy ETL~\cite{kargin2013lazy} demonstrates how metadata can be efficiently used for study-specific queries without actually constructing an entire data warehouse beforehand, by using files in SEED~\cite{ahern2007seed} standard format for seismological research. The hierarchical structure and metadata of SEED are similar to that of DICOM medical imaging data files that are accessed by the \projectName prototype. Thus, we note that while we prototype \projectName for medical research, the approach is also applicable to various research and application domains. 

LigDB~\cite{milchevski2015ligdb} is similar to \projectName as both focus on a query-based integration approach as opposed to having an entire data warehouse constructed as the first step, and it efficiently handles unstructured data with no schema. However, \projectName differs as it indeed has a scalable integrated data repository, and does not periodically evict the stored data, unlike LigDB. The incremental and selective integration and loading approach enables \projectName to load complex metadata faster than the current lazy ETL approaches.

\paragraph*{\textbf{Medical Research Data Integration:}}~
Leveraging Hadoop ecosystem for management and integration of medical data is not entirely new, and our choices are indeed motivated by previous work~\cite{lyu2015design}. However, the existing approaches fail to extend the scalable architecture offered by Hadoop and the other big data platforms to create an index to the unstructured integrated data, manage the data in-memory for quicker data manipulations, and share results and datasets efficiently with peers. \projectName attempts to address these shortcomings with its novel hybrid ETL approach and architecture, designed for reproducible scientific research.

\section{Conclusion}
\label{sec:conclusion}

\projectName is an on-demand data integration system with human-in-the-loop for scientific research. It selectively integrates and loads the data and metadata in a scalable integrated data repository. By implementing and evaluating \projectName for medical research data, we demonstrated the efficiency of the \projectName hybrid ETL process. We presented the \projectName data sharing approach to share scientific research datasets with minimal replication. 

We built our case on the reality that data sources are proliferating, and cross-disciplinary researches, such as medical data research, often require access and integration of datasets spanning across the multiple data sources on the Internet. We further presented how a human-driven selective ETL approach fits well for the reproducible scientific research. \projectName leverages the respective APIs offered by the data sources in accessing and loading the data while offering its RESTful APIs to access its integrated data repository. We further envisioned that various organizations with an \projectName instance would be able to collaborate and coordinate to construct and share the integrated datasets internally and between one another.

As a future work, we aim to deploy \projectName approach to consuming data from various scientific research data repositories such as EUDAT to find and integrate research data. Thus, we will be able to conduct a usability evaluation of \projectName based on various scientific research domains and data sources. We also propose to leverage the network proximity among the data sources and the \projectName instances for efficient data integration and sharing, in the future work. Thus, we aim to build virtual distributed data warehouses - data partially replicated and shared across various research institutes.











\bibliographystyle{spmpsci}      
\bibliography{references}   

%
%

\end{document}